\documentclass[pre,aps,10pt,twocolumn,floatfix,showpacs,superscriptaddress]{revtex4-1}
\usepackage{graphicx}
\usepackage{amsmath,amssymb}
\usepackage{natbib}
\usepackage{xcolor}

\def\nn{\nonumber}

\newcommand{\eref}[1]{Eq.~(\ref{#1})}%
\newcommand{\fref}[1]{Fig.~\ref{#1}} %
\newcommand{\sref}[1]{Sec.~\ref{#1}}%
\newcommand{\be}{\begin{equation}}
\newcommand{\ee}{\end{equation}}
\newcommand{\bea}{\begin{eqnarray}}
\newcommand{\eea}{\end{eqnarray}}

\begin{document}

\title{Ensemble inequivalence in the Blume-Emery-Griffiths model near a fourth order critical point}

\author{ V. V. Prasad} 
\affiliation{Department of Physics of Complex Systems, Weizmann Institute of Science, Rehovot 7610001, Israel}
\author{Alessandro Campa} 
\affiliation{National Center for Radiation Protection and Computational Physics, Istituto Superiore di Sanit\`{a},
Viale Regina Elena 299, 00161 Roma, Italy}

\author{David Mukamel} 
\affiliation{Department of Physics of Complex Systems, Weizmann Institute of Science, Rehovot 7610001, Israel} 
\author{Stefano Ruffo} 
\affiliation{SISSA, INFN and ISC-CNR, Via Bonomea 265, I-34136 Trieste, Italy}
\date{\today}

\begin{abstract}
The canonical phase diagram of the Blume-Emery-Griffiths (BEG) model with infinite-range interactions is known to exhibit a fourth
order critical point at some negative value of the bi-quadratic interaction $K<0$. Here we study the microcanonical phase diagram of this
model for $K<0$, extending previous studies which were restricted to positive $K$. A fourth order critical point is found to exist at
coupling parameters which are different from those of the canonical ensemble. The microcanonical phase diagram of the model close to the
fourth order critical point is studied in detail revealing some distinct features from the canonical counterpart.
\end{abstract}

\maketitle

\section{Introduction}

Long-range interacting systems have gained considerable attention, due to their unusual characteristics 
when compared with the widely studied systems with short range interactions~\cite{Campabook2014,dauxois2002dynamics}.
By long-range interacting systems, one refers to cases where the two-body interaction potential between  
degrees of freedom decays algebraically with the distance $r$ as $1/r^{d+\sigma}$, where $d$ is the spatial dimension and $\sigma \le 0$.
Such systems, for which the energy and other thermodynamic potentials are non-additive, are rather widely spread in nature, including
self-gravitating systems~($d=3$, $\sigma=-2$)~\cite{chavanis2002statistical,padmanabhan1990statistical}, interacting 
geophysical vortices~($d=2$ and logarithmic interaction)~\cite{chavanis2002statistical}, dipolar interactions in ferroelectrics and
ferromagnets~( $d=3$, $\sigma=0$)~\cite{landau1960electrodynamics}, and in plasmas~\cite{nicholson1992introduction} to name a few.
The case $\sigma = -d$ corresponds to infinite-range, mean-field interaction, which has conveniently been used to study various features
of long-range interacting systems. 

The non-additive nature of thermodynamic quantities in long-range systems makes them rather different from the more commonly studied systems
with short-range interactions, resulting in number of non-trivial features such as inequivalence of different
ensembles~\cite{Campa2009,Bouchet2010}. For example one finds that in these systems the entropy needs not be a concave function of
energy, which implies a negative specific heat in the microcanonical ensemble.
This is in contrast with what is obtained in the canonical ensemble. In addition at first order phase  transitions the temperature
displays a discontinuity in the microcanonical ensemble, a feature which is clearly absent in the canonical ensemble. Similar
features are found when grand-canonical and canonical ensembles are compared~\cite{Misawa2006}. The lack of additivity results in the
presence of non-convex domains in the parameter space of accessible thermodynamic variables and in breaking of
ergodicity~\cite{Borgonovi2004,MukamelPRL:2005}.
Various other interesting effects have been predicted in the relaxation of certain long-range systems to their final equilibrium state,
where the system approaches intermediate long lived `quasi-stationary states' before reaching
equilibrium~\cite{Lynden-Bell1967,Chavanis_1996,Latora1999,Yamaguchi2004}.

A simple paradigmatic model in which properties of systems with long-range interactions have been studied and ensemble inequivalence
has been demonstrated is the Blume-Emery-Griffiths (BEG) model, 
introduced to study the phase separation and transition to super fluidity in $\text{He}^3-\text{He}^4$ mixtures~\cite{BEG1971},  which was later generalized and used for studying generic two component fluid mixtures~\cite{Mukamel-Blume1974,Krinsky:1975}.
This is a spin-1 lattice model with both bilinear and biquadratic spin-spin interactions. In the case of infinite-range interactions,
where every spin interacts with every other spin with the same coupling constants ~($\sigma=-d$), the Hamiltonian of the model can be
represented as:
\be
H=\Delta\displaystyle\sum_{i=1}^{N}S_i^2-\frac{J}{2N}\displaystyle\left(\sum_{i=1}^{N}S_i\right)^2-\frac{K}{2N}\displaystyle\left(\sum_{i=1}^{N}S_i^2\right)^2,
\label{hamiltonian}
\ee
where each spin $S_i$ takes one of the values $\{-1,0,1\}$. The parameter $\Delta$ controls the energy difference between the
ferromagnetic ($S_i=\pm1$) and the paramagnetic  ($S_i=0$) states, $J>0$ is a ferromagnetic coupling and $K$ is a biquadratic coupling
which could have either sign. Even though each spin interacts with every other spin, the scaling $J/N$ and $K/N$ makes the energy
extensive (although not additive). Without loss of generality one may take $J=1$.

\begin{center}
\begin{figure}
\includegraphics[width=0.48\textwidth, angle=0]{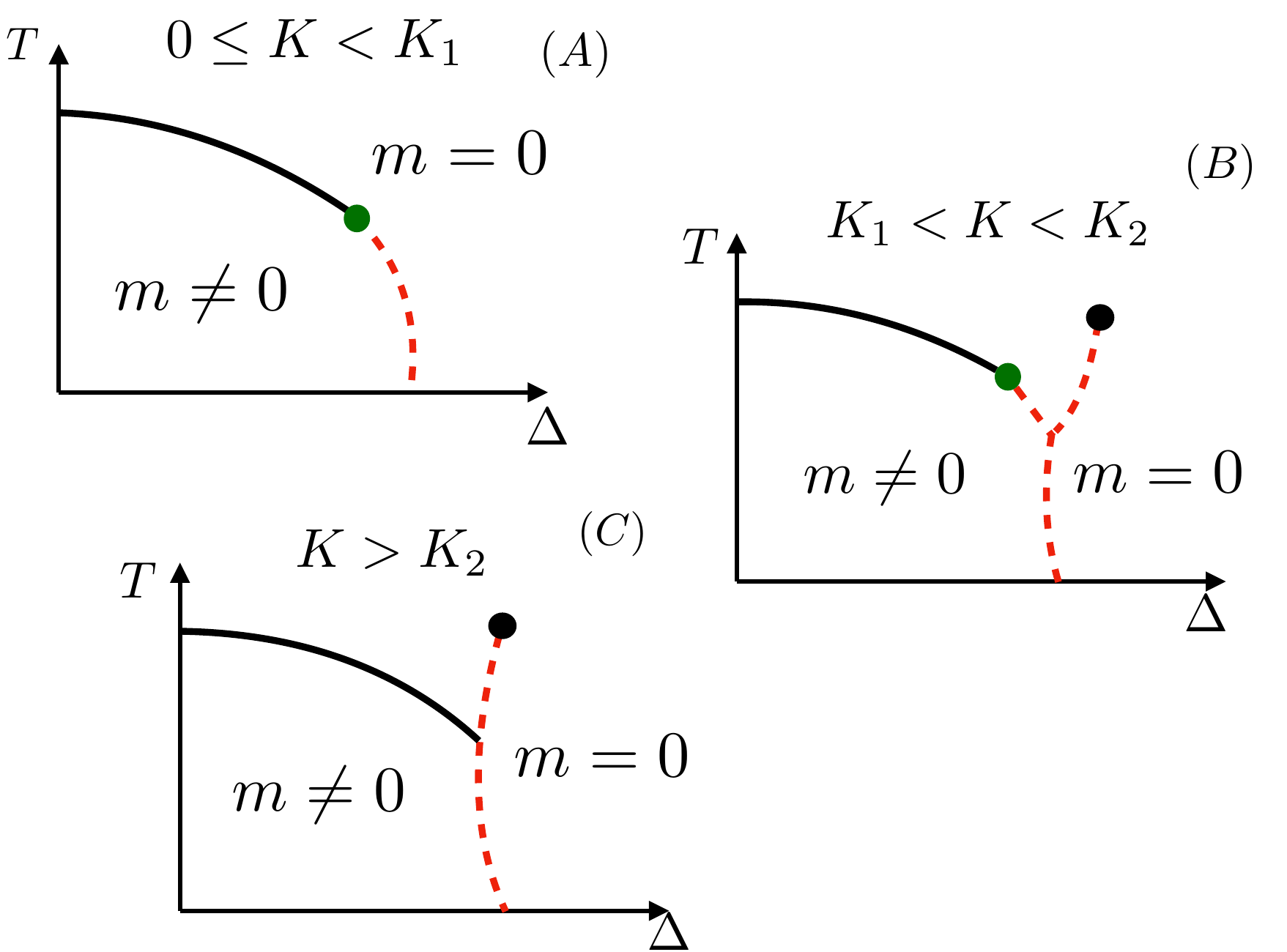}
\caption{(color online) Schematic plot showing the canonical phase diagram in the $(\Delta,T)$ plane for different domains of
$K\geq 0$. The plot~(A) corresponds to the range $0<K<K_1\approx2.775$, where the transition line separating the
ferromagnetic~(ordered, $m\not=0$~[see~\eref{order parameter} for the definition of $m$]) and
paramagnetic~(disordered, $m=0$) phase, is composed of continuous~(solid black) and first order~(dashed red) line segments meeting at
the tricritical point~(green dot). Plot~(B) is for $2.775\approx K_1<K<K_2\approx3.801$, where additional to the features described
for $K<K_1$, the first order line branches at a triple point into the disordered phase, indicating a transition between two disordered
phases with different quadrupole moment. This branch terminates at a critical point (black dot). Plot~(C) is for  $K>K_2$, where
the second order and first order lines join at a critical end point rather than at a tricritical one.}
\label{fig001}
\end{figure}
\end{center}

The canonical phase diagram of the model (\ref{hamiltonian}) has been shown to display unique features at different domains of model
parameters~\cite{BEG1971,Mukamel-Blume1974,Krinsky:1975,Lajzerowicz1975,Hoston:1991}. 
For fixed $K \ge 0$ the $(\Delta,T)$ phase diagram exhibits a ferromagnetic ordered phase at small values of $T$ and 
$\Delta$, and a paramagnetic disordered phase, otherwise. However some qualitative features of the phase diagram are modified as
$K$ increases, as shown in \fref{fig001}. At small $K$ the transition line between the two phases changes character from
continuous~(solid black) to first order~(dashed red) at a tricritical point~(green dot). At higher values
of $K$~\cite{Hovhannisyan2017}, another first order line emerges, separating two disordered phases. The two first order lines meet
at a triple point where the two disordered phases coexist with the ordered one~[See \fref {fig001}~(B)]. For even larger $K$ values, the
tricritical point becomes a critical end point~[See \fref {fig001}~(C)] and the continuous branch of the transition line terminates
at the intersection with the first order line. 

The canonical phase diagram for $K<0$ has also been addressed~\cite{Krinsky:1975,Hoston:1991}. It has been shown that while at
small $K$ the phase diagram is qualitatively similar to the $K=0$ one (with first and second order lines joining at a tricritical point),
at some particular value of $K$ the tricritical point becomes a fourth order one. Beyond that value, the phase diagram becomes rather
different from that of positive $K$. While the second order line terminates at the first order one at a critical end point
(as in the $K>0$ regime), the first order line enters into the ordered phase separating two distinct ferromagnetically ordered
phases (see \fref{fig010}). This phase diagram holds, schematically, for a range of values of negative $K$. 

The microcanonical phase diagram was studied for the model for $K=0$~\cite{Barre2001} and for $K>0$~\cite{Hovhannisyan2017},
illustrating the inequivalence between the two ensembles. It has been demonstrated that while the two ensembles have a common critical
line at small values of $K$, the two ensembles yield distinct phase diagrams in the region where the canonical transition is first order.
In particular the microcanonical tricritical point is located at a different point in the phase space of the model. Detailed studies of
the phase diagram in the vicinity of the tricritical point show that the microcanonical first order line does not coincide with its
canonical counterpart, and that it involves temperature discontinuity, which is of course missing in the canonical treatment. Furthermore,
analysis of large positive values of $K$ reveals a wealth of different features in the phase diagrams of the two
ensembles ~\cite{Hovhannisyan2017}.

In the present paper we extend the study of the microcanonical phase diagram of the infinite range BEG model to negative values of the
parameter $K$ where a fourth order critical point has been found in the canonical phase diagram. As is usually the case, a high order
critical point determines the topological features of the phase diagram around it and the way the various phase transition manifolds
join together. These topological features tend to persist in quite a broad range of the model parameters, making a study of this point
of particular interest. The fact that the canonical phase diagram of this model exhibits a fourth order
critical point at some negative value of $K$ suggests that such a point may also exist in the microcanonical phase diagram as well,
which would enable one to make a detailed comparison between the phase diagrams of the model obtained in the two ensembles. We find that
indeed the microcanonical phase diagram exhibits a fourth order critical point at negative $K$, located at a different point in phase
space as compared with the canonical one. We analyze the global features of the microcanonical phase diagram and discuss the way the
inequivalence between the two ensembles is manifested in this parameter region.

The rest of the paper is as follows: A brief outline of the analysis of BEG model in the canonical ensemble is presented
in \sref{canonical analysis}, which allows us to display the  phase diagram in the relevant region  of the parameter space. The
analysis is carried out for $K<0$, for which the $4$-th order transition point is present. In \sref{microcanonical}, the microcanonical
analysis of the model is presented and the fourth order point in this ensemble is identified. In \sref{kplane}, we discuss in detail
the microcanonical phase diagram around the fourth order critical point. Concluding remarks are given in \sref{conclusions}.

\section{Canonical phase diagram}
\label{canonical analysis}
In the following, we briefly outline the derivation of the canonical phase diagram for negative $K$, where a fourth order critical
point is found to be present. Note that in non-additive systems, such as the one considered in this paper, the canonical ensemble cannot be simply derived from the microcanonical one. For a discussion of this point see ~\cite{baldovin2018,rocha2018}. Here we consider the partition function of the system,
\be
Z(\beta=1/T,N)=\displaystyle\sum_{\{S_i\}}e^{-\beta H},
\label{partition function}
\ee
where $H$ is as given in \eref{hamiltonian}, with  the Boltzmann constant $k_B=1$. 
Let
\bea
m=\sum_{i=1}^NS_i/N~~\text{and}~~~q=\sum_{i=1}^NS_i^2/N,
\label{order parameter}
\eea
be the magnetization $m$ and quadrupole moment $q$ order parameters, respectively. 
The partition function can be calculated by converting the right hand side of \eref{partition function} into an integral using the
Hubbard-Stratonovich transformation. Making use of the gaussian identity,
\bea
e^{a b^2}=\sqrt{\frac{|a|}{\pi}}\displaystyle\int_{-\infty}^\infty dx~ e^{-|a|x^2+2sabx}
\eea
where $s=1$ for $a>0$ and $s=i$ (the imaginary unit) for $a<0$, one can represent the partition function as
\bea
Z(\beta,N)=&&\frac{N\beta \sqrt{-K}}{2 \pi}\displaystyle\sum_{\{S_i\}}e^{-\beta N\Delta q}\times\nn\\
\displaystyle\int\limits_{-\infty}^{\infty}\displaystyle\int\limits_{-\infty}^{\infty}&&dx~ dy~ e^{{-\frac{\beta N}{2}x^2+
\frac{\beta NK}{2}y^2+\beta N m x+\beta N K q iy}},
\eea
where $x$ and $y$ are the corresponding auxiliary fields. We point out that in this paper we are considering only
positive temperatures, although this model, where the energy is upper bounded, allows also negative temperatures in the microcanonical ensemble~\cite{Hovhannisyan2017}.
Therefore in the following it is always $\beta >0$.

Performing the sum over $\{S_i\}$ results in
\be
Z(\beta,N)=\frac{N\beta \sqrt{-K}}{2 \pi}
\displaystyle\int\limits_{-\infty}^{\infty}\displaystyle\int\limits_{-\infty}^{\infty}dx ~dy e^{-\beta N\tilde{f}(\beta,x,y)},
\label{partition function 2}
\ee
where,
\bea
\beta\tilde{f}(\beta,x,y)=&&\frac{\beta}{2}(x^2-K y^2)\nonumber\\&&-\ln\left[1+2 e^{-\beta\Delta+\beta K iy}\cosh(\beta x) \right].
\label{partition function exponent}
\eea
The integration can be done using a saddle point analysis in terms of the variables $x$ and $y$. Note that the values  $x$ and $iy$ which
minimize $\beta\tilde{f}(\beta,x,y)$, correspond respectively to the equilibrium magnetization $m$, and the quadrupole moment $q$, where
the minimizing value of $y$ is purely imaginary. At the saddle point one obtains
 \bea
\label{eq magentisation}
x=\frac{2 \sinh \beta x}{e^{\beta\Delta-i\beta K y}+2\cosh \beta x},\\
iy=\frac{2 \cosh \beta x}{e^{\beta\Delta-i\beta K y}+2\cosh \beta x}.
\label{eq quadrupole moment}
\eea
Furthermore, for a non-zero magnetization (which corresponds to  $x\not=0$) the above relations also lead to the expression:
\bea
iy=x\coth \beta x.
\label{q fn m}
\eea
One will find these relations to be useful when characterising the phase diagram, as explained below.

\begin{figure}
\includegraphics[width=0.45\textwidth, angle=0]{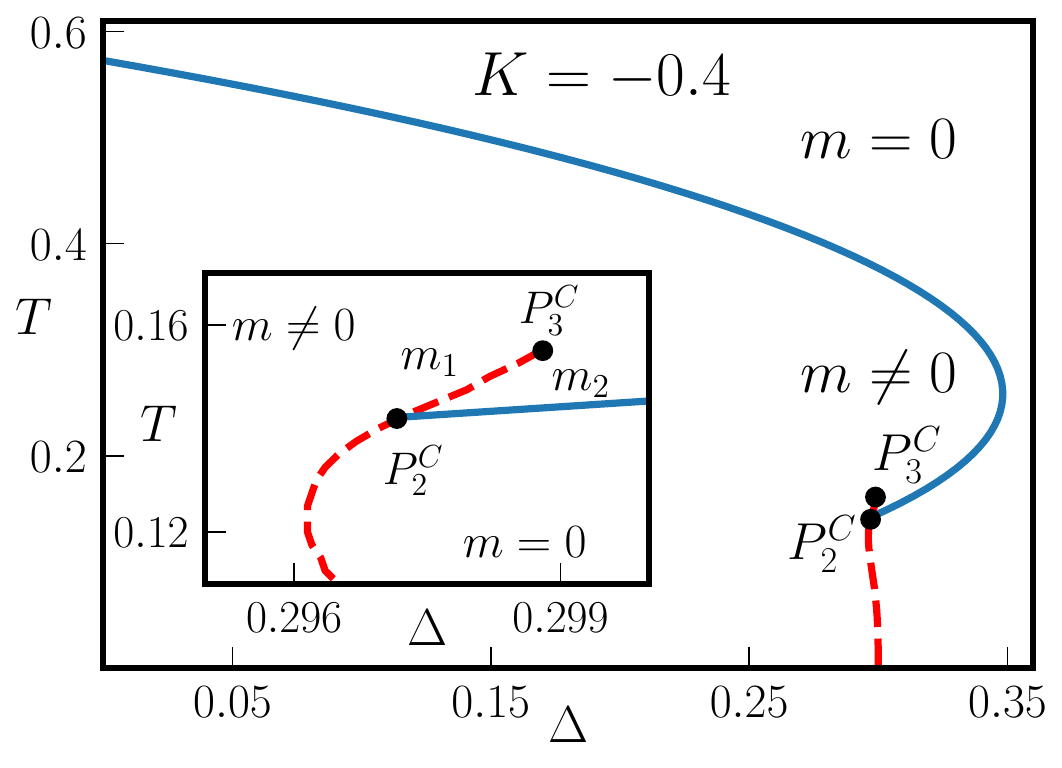}
\caption{(color online) The canonical phase diagram in the $(\Delta,T)$ plane for $K=-0.4$. Here, the continuous transition line~(solid blue) 
meets the first order line~(dashed red) at a critical end point denoted by $P_2^C~(\Delta \approx 0.2972,~T\approx 0.1412)$. The first order line
extends into the ordered phase, where it marks a transition between two ordered phases with different values of $m$ and $q$, and
terminates at the critical point $P_3^C~(\Delta\approx 0.29873,~T \approx 0.1544)$. The zoomed in phase diagram close to this
transition is plotted in the inset.}
\label{fig010}
\end{figure}

To obtain the critical line one expresses $y$ in terms of $x$ using \eref{q fn m} and expands
the free energy $\tilde{f}(\beta,x,y)$ about the paramagnetic solution  $x=0$ and $iy=1/\beta$~[see Eqs. (\ref{eq magentisation})
and (\ref{q fn m})] in powers of $x$,  
\bea
\beta f(\beta,x,y (x))=&& \beta f_o+A_c x^2+B_c x^4+C_cx^6+D_cx^8 \cdots
\nonumber \\ &&
\eea
where  $f_0$ is the free energy value at $x=0$, 
\bea
A_c=\frac{\beta (3 + 2K)}{6}\left[1-\frac{2\beta}{2+e^{\beta\Delta-K}}\right]
\eea 
and $B_c, C_c$ and $D_c$ are given by more complicated expressions of $\beta, \Delta$ and $K$ which are not displayed here. 
The critical surface is obtained at $A_c=0$, yielding 
\bea
\beta=1+\frac{1}{2}e^{\beta\Delta-K}.
\label{critical line}
\eea
The critical surface represents a locally stable solution as long as $B_c$ is positive. On the critical surface (\eref{critical line})
the coefficient $B_c$ takes the form    
\bea
B_c=\frac{\beta^2}{72}(2K+3)[(2K+3)-\beta(2K+1)].
\label{tricritical point}
\eea
 Considering $K > -0.5$, the region where the fourth order critical point is located, the critical surface is stable
for
$(3+2K)/(1+2K)>\beta$ and it terminates on a tricritical line obtained at $B_c=0$, namely, at
\bea
\beta=\frac{3+2K}{1+2K}.
\label{tricritical line}
\eea
Equations (\ref{critical line}) and (\ref{tricritical line}) thus yield the tricritical line in the 3-dimensional space spanned by
$(T,\Delta, K)$. This line is stable as long as $C_c>0$ and it terminates at a fourth order critical point at which $C_c=0$.
On the tricritical line, where $A_c=B_c=0$, $C_c$ takes the form    
\bea
C_c=\frac{\beta^5}{1620(\beta-1)^2}(9+2\beta-\beta^2).
\label{C c along tricritical line}
\eea
It vanishes at $\beta^2-2\beta-9=0$. This equation, together with (\ref{critical line}) and (\ref{tricritical line}), yields the fourth
order critical point 
\bea
T^*=&&(1+\sqrt{10})^{-1} \approx 0.2402,\nn\\K^*=&&(3T^*-1)/[2(1-T^*)] \approx -0.1838,
\label{canonical fourth order}\\
\Delta^*=&&T^*\left(K^*+\ln\left[\frac{2(1-T^*)}{T^*}\right]\right) \approx 0.399.\nonumber
\eea

In order to complete the phase diagram one has to find the global minimum of the free energy $\tilde{f}(\beta,x,y)$ which is done
numerically. In \fref{fig010} the phase diagram in the  $(\Delta,T)$ plane for fixed $K<K^*$ is displayed. One finds in the figure
both ordered and disordered phases separated by continuous (solid) and first order transition (dashed) lines.
The critical line terminates on the first
order surface at a critical end point denoted by $P_2^C$. The first order line is composed of two segments, one separating a
paramagnetic from a ferromagnetic phase, and the other separating two magnetically ordered phases $m_1$ and $m_2$, with $m_1\not=m_2$.
This segment terminates at a critical point, labelled as~$P_3^C$. The two magnetically ordered phases are characterized also
by different quadrupole moments $q$. This can be seen from Eq. (\ref{q fn m}) (we recall that at the minimum of the free energy, $x$ and $iy$  correspond to the
equilibrium magnetization and quadrupole moment, respectively): at given $\beta$, a jump in $x$ implies a jump in $y$.

\section{Micro-canonical Analysis}
\label{microcanonical}

In order to analyze the phase diagram of the model within the microcanonical ensemble we note that the energy of any microscopic
configuration can be expressed in terms of only two parameters: the  total number of up-spins $N_+ $ and total number of down
spins $N_-$. The number of spins taking the value $S=0$, $N_0$, is simply related to $N_+$ and $N_-$  by $N_++N_-+N_0=N$. The
energy (\ref{hamiltonian}), is thus given by 
\be
E=\Delta Q -\frac{1}{2N}M^2-\frac{K}{2N}Q^2 
\label{microcanonical total energy}
\ee
where $M=N_+-N_-$ and $Q=N_++N_-$, which are the magnetic and quadrupole moments respectively. To calculate the entropy associated
with the macroscopic state defined by $M$ and $Q$, one has to enumerate the possible microscopic configurations $W$ specified by
the values of  $N_+,~N_-$ and $N_0$. This is given by
\be
W=\frac{N!}{N_+!N_-!N_0!}~~.
\ee
\begin{figure*}[t]
\includegraphics[width=0.95\textwidth, angle=0]{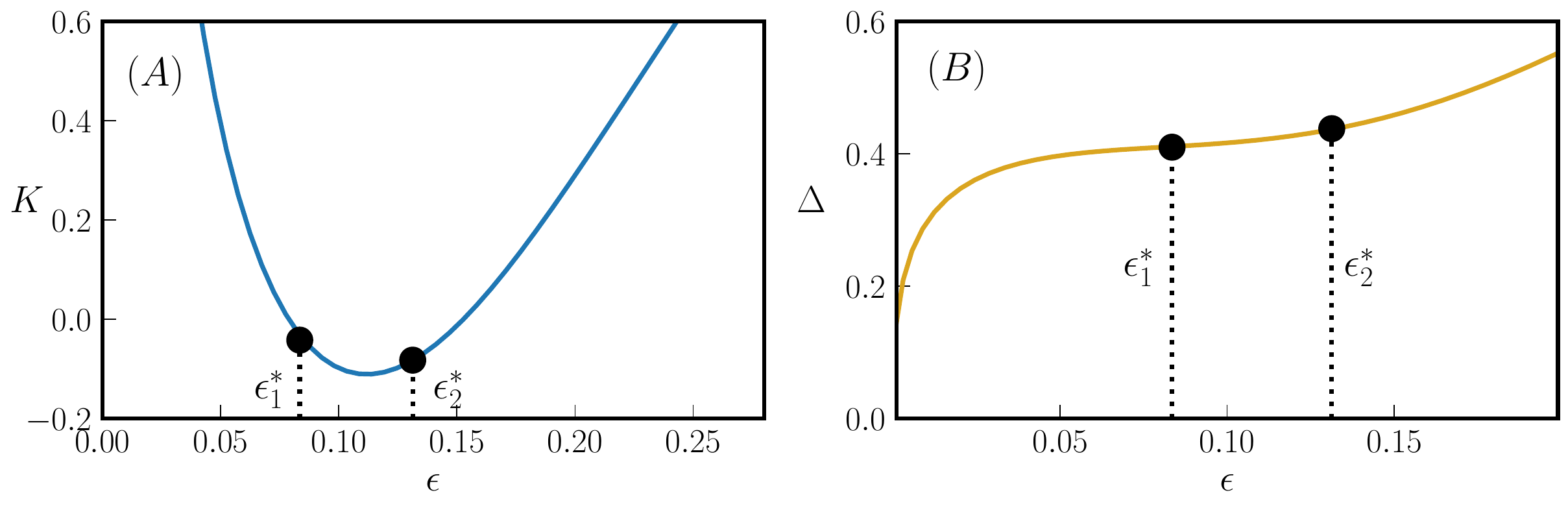}
\caption{(color online) The tricritical line obtained from the micro-canonical analysis by solving for $A_m=B_m=0$
[see \eref{entropy landau expansion}], plotted both in the $(\epsilon,K)$ plane~[Panel~(A)] and
$(\epsilon,\Delta)$ plane~[Panel~(B)]. Solutions for the fourth order critical points,
$\epsilon^*_{1}\approx0.0835$ and $\epsilon^*_{2}\approx0.1313$, are indicated by black dots.}
\label{fig01a}
\end{figure*}
In the large $N$ limit the entropy $S=\ln W$ is 
\bea
S=&-&N[(1-q)\ln(1-q)+\frac{1}{2}(q+m)\ln(q+m)\nn\\&+&\frac{1}{2}(q-m)\ln(q-m)-q\ln 2],
\label{entropy-exact}
\eea
where $m=M/N$ and $q=Q/N$ are the single site magnetic and quadrupole  moments respectively.
The entropy at equilibrium  can now be obtained by maximizing \eref{entropy-exact} at a fixed energy value $E$. 

Expressing the single site energy $\epsilon=E/N$ in terms of the single site macroscopic quantities $m$ and $q$,
Eq. (\ref{microcanonical total energy}) becomes
\be
q^2-2\frac{\Delta}{K}q+\frac{2\epsilon}{K}+\frac{m^2}{K}=0.
\ee
This allows a solution for  $q$ in terms of $m$ and $\epsilon$: 
\be
\label{q_fun_eps_m}
q_\pm=\frac{\Delta}{K}\pm\sqrt{\left(\frac{\Delta}{K}\right)^2-\frac{2\epsilon}{K}-\frac{m^2}{K} }~.
\ee
For given values of the parameters $\Delta$ and $K$ and of the magnetization $m$, the energy $\epsilon$ must be in a range such that
the expression under square root is not negative. For $K<0$, the only acceptable solution is $q_+$, since $q_-$ is negative.  
Substituting the solution for $q_+$ in the expression for the entropy~(\ref{entropy-exact}), one obtains the single site entropy
$S/N=\tilde{s}_+(\epsilon,m)$ as a function of $\epsilon$ and $m$. The equilibrium entropy corresponds to the global maximum
of $\tilde{s}_+(\epsilon,m)$ as a function of $m$, i.e., $s_+(\epsilon)=\max_m[\tilde{s}_+(\epsilon,m)]$.

In order to find the critical and multicritical surfaces of the phase diagram we expand the entropy $\tilde{s}_+(\epsilon,m)$ around
the paramagnetic phase $m=0$. The expansion takes the form 
\bea
\tilde{s}_+(\epsilon, m)=&&s_0+A_m m^2+B_m m^4+C_m m^6+\nonumber\\&&D_m m^8+O(m^{10})+\cdots
\label{entropy landau expansion}
\eea
where $s_0$ is the zero magnetization entropy: 
\be
s_0=-(1-z_+)\ln(1-z_+)-z_+\ln z_++z_+\ln 2
\ee
with $z_+=q_+(m=0)$. The expansion coefficients are given by
{\small\bea
\label{coefficients microcanonical}
A_m&=&- a \ln \frac{2(1-z_+)}{z_+}-\frac{1}{2z_+},\nn\\
B_m&=&- Ka^3 \ln \frac{2(1-z_+)}{z_+}-\frac{a^2}{2z_+(1-z_+)}-\frac{a}{2 z_+^2}-\frac{1}{12 z_+^3},\nn\\
 C_m&=&- 2 K^2 a^5\ln \frac{2(1-z_+)}{z_+}-\frac{Ka^4}{z_+(1-z_+)}\nn\\&&+\frac{a^3(2 z_+-1)}{6 z_+^2(1-z_+)^2}-\frac{K a^3}{2 z_+^2}-
 \frac{a^2}{2 z_+^3}-\frac{a}{4 z_+^4}-\frac{1}{30z_+^5},\nn\\
 \label{defdm}
 D_m&=&- 5 K^3 a^7\ln \frac{2(1-z_+)}{z_+}-\frac{5K^2a^6}{2z_+(1-z_+)}\nn\\&&+\frac{Ka^5(2 z_+-1)}{2 z_+^2(1-z_+)^2}\nn
 -\frac{a^4(1-3 z_++3z_+^2)}{12 z_+^3(1-z_+)^3}
-\frac{K^2 a^5}{ z_+^2}\nn\\&&-\frac{Ka^4}{z_+^3}-\frac{a^3}{2 z_+^4}-\frac{K a^3}{4 z_+^4}
-\frac{a^2}{2z_+^5}-\frac{a}{6z_+^6}-\frac{1}{56z_+^7},
\eea}where the subscript $m$ denotes the micro-canonical coefficients and
	\be
a=~\text{sgn}(K)\left(4\Delta^2 - 8K\epsilon\right)^{-\frac{1}{2}}.
\label{a}
\ee

The critical surface in the $(\epsilon, \Delta, K)$ space is obtained at $A_m=0$ with $B_m<0$. 
To obtain the expression giving $A_m=0$ we start from the microcanonical inverse temperature,
given by
\bea
\beta=\frac{\partial \tilde{s}_+}{\partial\epsilon}=\ln\left[\frac{2(1-q_+)}{\sqrt{q_+^2-m^2}}\right]\frac{\partial q_+}{\partial\epsilon},
\eea
where $m$ takes the value which maximizes $\tilde{s}_+(\epsilon,m)$. On the critical line, where $m=0$, this expression becomes
\be
\label{betaoncrit}
\beta = \frac{\partial s_0}{\partial \epsilon} = -2a \ln \frac{2(1-z_+)}{z_+} \, .
\ee
Substituting this equation in $A_m=0$ one obtains $z_+ = 1/\beta$. Inserting this into Eq. (\ref{betaoncrit}) we have
$\beta=\frac{1}{2}e^{-\frac{\beta}{2a}}+1$. On the other hand, from the definition of $a$ given in Eq. (\ref{a})
we obtain $\beta/(2a) = K -\beta \Delta$. So at the end $A_m=0$ is expressed by
\bea
\beta=\frac{1}{2}\exp[{\beta \Delta-K}]+1,
\label{MC critical line}
\eea
Note that the expression of the critical surface is the same as the one obtained for the canonical ensemble, as
expected~\cite{Barre2001,Campa2009}.

Similarly the tricritical line marking the termination of the critical surface, is obtained at $A_m=B_m=0$, with $C_m<0$. These
equations can be solved and the tricritical line in the $(\epsilon,~\Delta,~K)$ space can be expressed in terms of the parameter
$\beta$ as   

\bea
K(\beta)&=&\frac{\beta}{\beta-1}-2\ln(2\beta-2)+\frac{2}{3}[\ln(2\beta-2)]^2,\nonumber\\
\Delta(\beta)&=&[K(\beta)+\ln(2\beta-2)]\beta^{-1},\\
\epsilon(\beta)&=&\frac{\Delta(\beta)}{2\beta}+\frac{\ln(2\beta-2)}{2\beta^2}.\nonumber
\eea

In \fref{fig01a} we represent the tricritical line by plotting $K$, [Panel (A)] and $\Delta$, [Panel (B)] as a function of $\epsilon$.

\begin{figure}
\includegraphics[width=0.47\textwidth, angle=0]{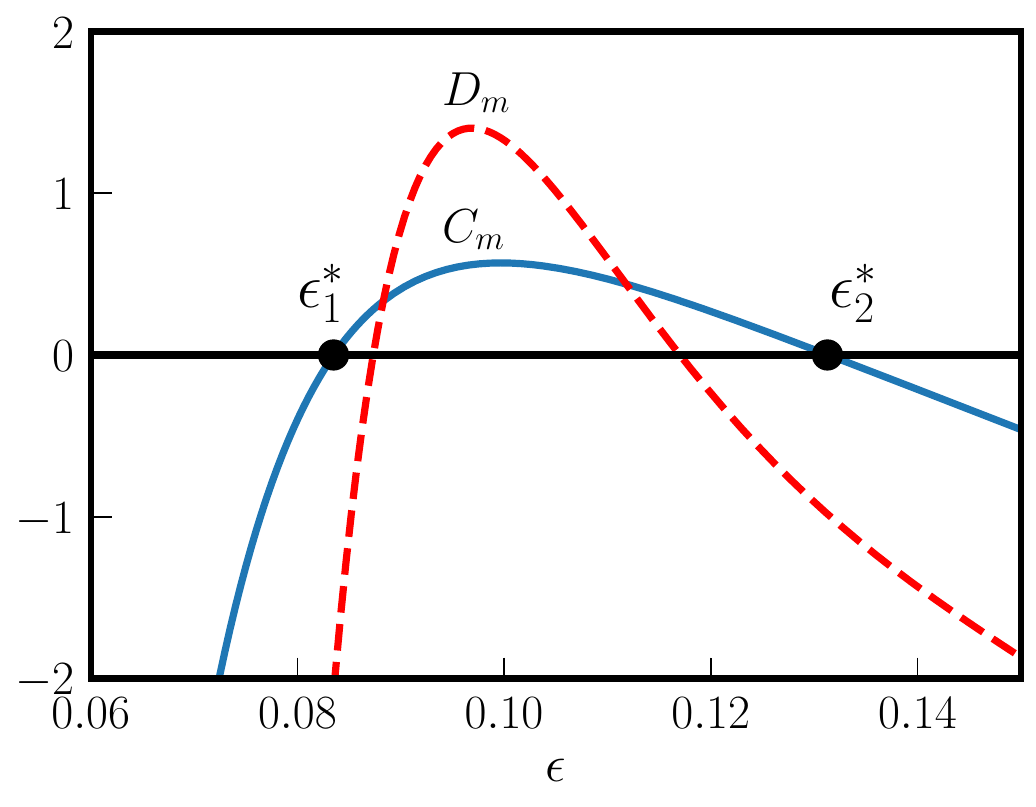}
\caption{The values of the coefficients $C_m$ and $D_m$ [see~\eref{entropy landau expansion}], plotted as a function of
the energy per particle $\epsilon$, for the parameter values corresponding to the tricritical line ($A_m=B_m=0$). 
The fourth order critical points can be read out from the figure, corresponding to $C_m=0$ and $D_m<0$, with values
$\epsilon^*_{1}\approx0.0835$ and $\epsilon^*_{2}\approx0.1313$.}
\label{fig01}
\end{figure}
 
The tricritical line terminates at the fourth order critical point which is obtained at $A_m=B_m=C_m=0$ with $D_m<0$. The three
constraints yield a point in the parameter space. To find the solution to these equations we plot (\fref{fig01}) the coefficients
$C_m$ and $D_m$ as a function of $\epsilon$, along the tricritical line. One can see that there are two solutions corresponding to two
energy values, $\epsilon_1^*\approx0.0835$ and $\epsilon_2^*\approx0.1313$ at which the sixth order coefficient  $C_m$ in  the expansion
vanishes. At both solutions the 8-th order coefficient is $D_m<0$ indicating that both are locally stable solutions. We will see below
that the only solution which corresponds to a global maximum of the entropy is $\epsilon_2^*$. The other solution
is preempted by a global maximum away from $m=0$. Thus the fourth order critical point of the microcanonical ensemble takes place at
\begin{equation}
\epsilon_2^*\approx 0.1313, ~~ \Delta_2^*\approx 0.4369,   ~~K_2^*\approx-0.0828,
\end{equation}
which corresponds to $T^*\approx0.2924$. Comparing these values with the fourth order point found from the canonical
calculation~(\ref{canonical fourth order}) shows that that the two differ from each other.

\begin{center}
\begin{figure}
\includegraphics[width=0.48\textwidth, angle=0]{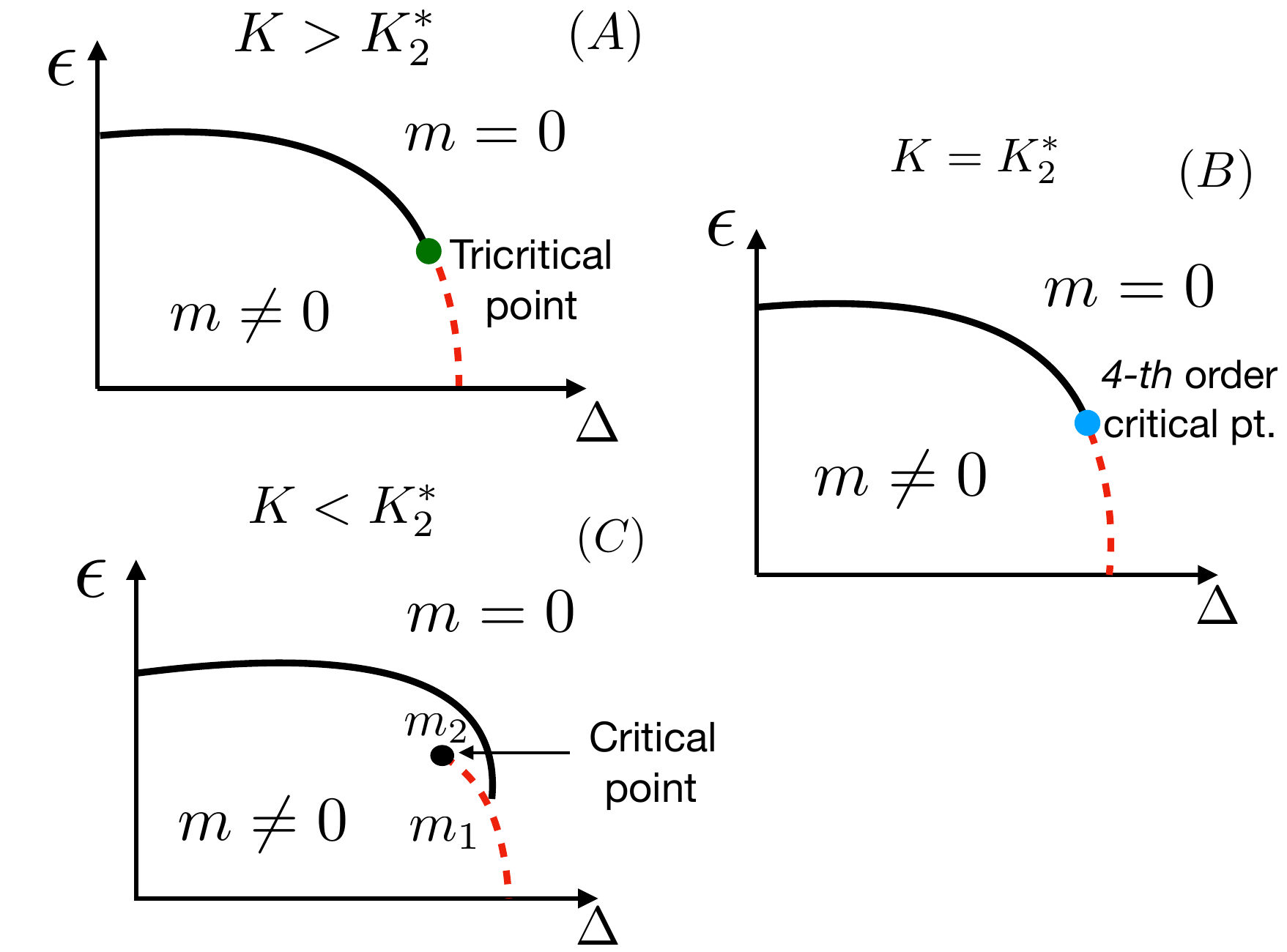}
\caption{Schematic microcanonical phase diagram near to the fourth order critical point.  The panels show the phase
diagrams in the ($\Delta,\epsilon$) plane for fixed $K$ values. Panel~(A) displays a $K>K^*_2$ plane where the ordered phase $(m\not=0)$ and
the disordered phase~$(m=0)$ are separated by two transition line segments: a continuous~(solid) and a first order~(dashed) transition line 
that merge at the tricritical point. Panel~(B) corresponds to $K=K^*_2$. It is similar to panel~(A), however the two segments of the
transition line join at a fourth order point rather than a tricritical one. In panel~(C), which corresponds to $K<K^*_2$ the two
segments of the transition line join at a critical end point. The first order transition line extends into the ordered phase indicating
transitions between two different ordered phases with magnetization values $m_1$ and $m_2$.} 
\label{fig01c}
\end{figure}
\end{center}

To complete the phase diagram one has to determine the first order surfaces of the model. This is done by numerically finding the
global maximum of the entropy. Before analyzing the detailed phase diagram near the fourth order critical point, which will be presented
in the next section, let us display the global features of the phase diagram. A schematic phase diagram in the ($ \Delta,\epsilon$) plane
for some values of $K$ is given in \fref{fig01c}. For $K>K^*_2$, the phase diagram consists of a transition line from a paramagnetic to
a ferromagnetically ordered phase which changes character from second order to first order at a tricritical point. At $K=K^*_2$ the
tricritical point becomes a fourth order point, and for $K<K^*_2$ the first order line extends into the magnetically ordered phase, 
indicating a transition between two ordered phases, and
the second order line terminates at a critical end point. Also in the microcanonical ensemble, the two ordered phases between which
a first order transition takes place are characterized by different magnetization and a different quadrupole moment. This can be seen
from Eq. (\ref{q_fun_eps_m}): at given $\epsilon$, a jump in $m$ implies a jump in $q_+$.  
It is evident that the microcanonical $(\Delta,\epsilon)$ phase diagram is
qualitatively similar to the canonical $(\Delta,T)$ as discussed in the preceding section. 
In particular, the qualitative features of the canonical ($\Delta,T$) phase diagram for $K$ larger, equal and smaller than $K^*$, are, respectively, similar to those shown in \fref{fig01c} concerning
the ($\Delta,\epsilon$) phase diagram for $K$ larger, equal and smaller than $K^*_2$.
In the next section we consider the detailed
microcanonical phase diagram near the fourth order critical point and present it in the $(T, \Delta, K)$ space, where the comparison with
the canonical phase diagram reveals the inequivalence between the two ensembles.

\begin{figure}
\centering
\includegraphics[width=0.46\textwidth]{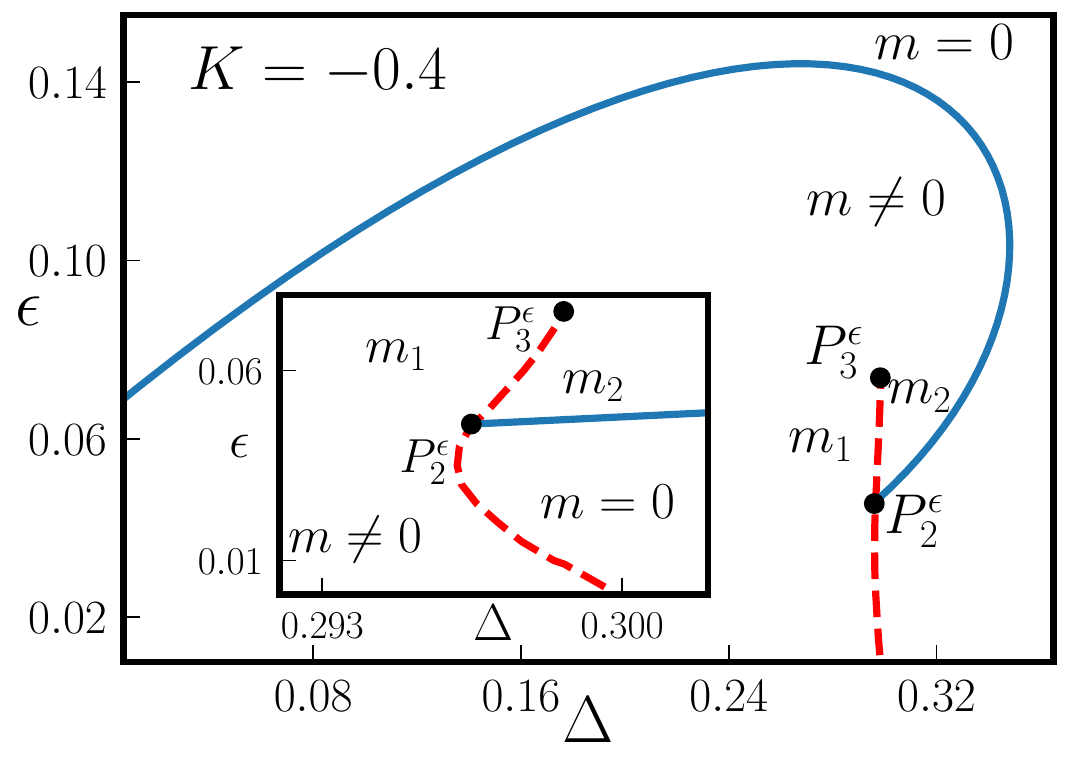}
\caption{(color online) The microcanonical phase diagrams in the $(\Delta,\epsilon)$ plane for a fixed value of $K=-0.4$. The locus of
continuous transition points in the plane is plotted as a solid line~(in blue) and the first order locus with dashed line~(in red).
The first order transition line is seen to extend into the ordered phase, indicating a transition between two ordered phases. The
continuous transition line terminates  at $P_2^\epsilon~(\Delta\approx0.2966,~\epsilon\approx0.0459)$. The critical point
$P_3^\epsilon~(\Delta\approx0.2987,~\epsilon\approx0.075)$  marks the termination of the first order line in the ordered phase.
The inset shows a zoomed in plot near the region in the phase diagram close to the first order transition, indicating the
re-entrant behavior in the ordered phase, as the energy is varied at fixed $\Delta$.}
\label{fig04}
\end{figure}

\begin{center}
\begin{figure*}[t]
\includegraphics[width=0.89\textwidth, angle=0]{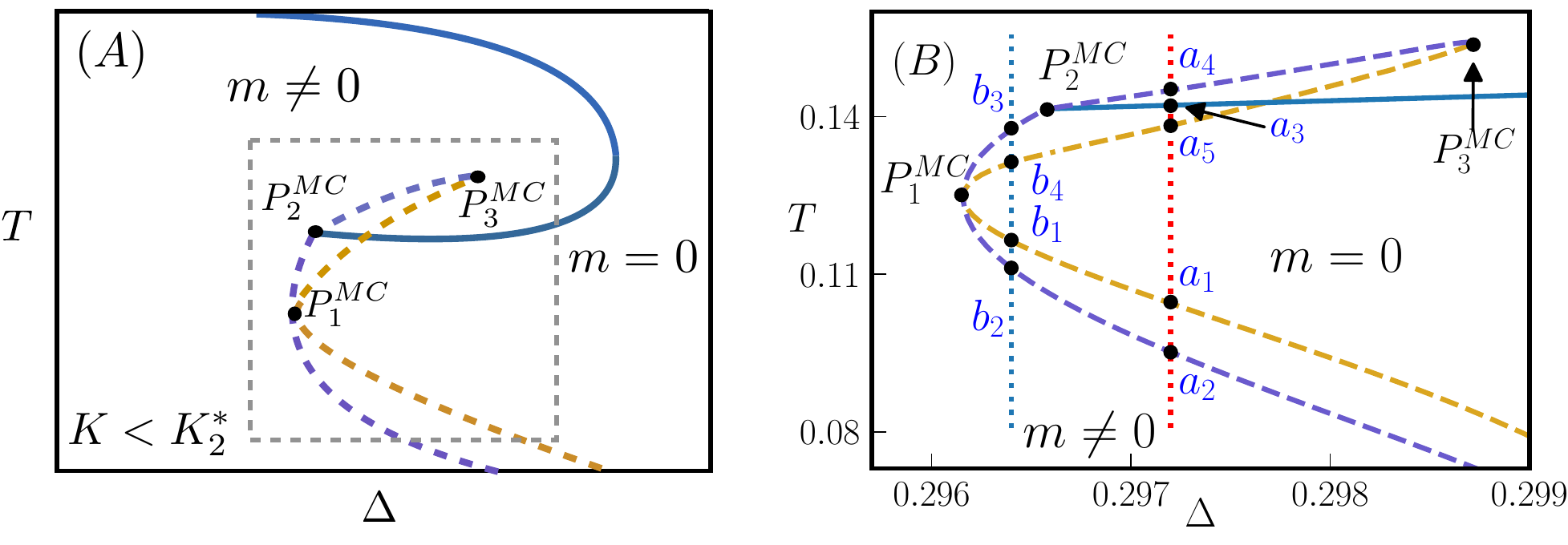}
\caption{(color online) Panel (A): Schematic phase diagram in  the $(\Delta,T)$ parameter space with fixed value of $K<K^*_2$ to show the 
relevant region in the plane where the transition is observed. In panel (B), the non-schematic phase diagram is plotted for $K=-0.4$;
this is an enlarged version of a section of the phase diagram [specified by the grey dashed square displayed in panel (A)]. In both
plots, the continuous transition line is shown as a solid curve and the first order transition lines as dashed curves. The label
$P_1^{MC}$ denotes the point in the parameter space, where the discontinuity in temperature vanishes. Furthermore, $P_2^{MC}$
represents the point at which the continuous transition line joins the first order line, and $P_3^{MC}$ is the critical point, where
the first order line terminates. The two vertical lines in panel (B) indicate fixed $\Delta$ lines, $\Delta=0.2972$ (red dotted line)
and  $\Delta=0.2964$ (blue dotted line)  along which the temperature profile is plotted  as a function of $\epsilon$
in \fref{fig06}. The labels \{$a_i$\} and \{$b_i$\} correspond to the various transition points along the respective lines.}
\label{fig05}
\end{figure*}
\end{center}

\begin{center}
\begin{figure}
\includegraphics[width=0.415\textwidth, angle=0]{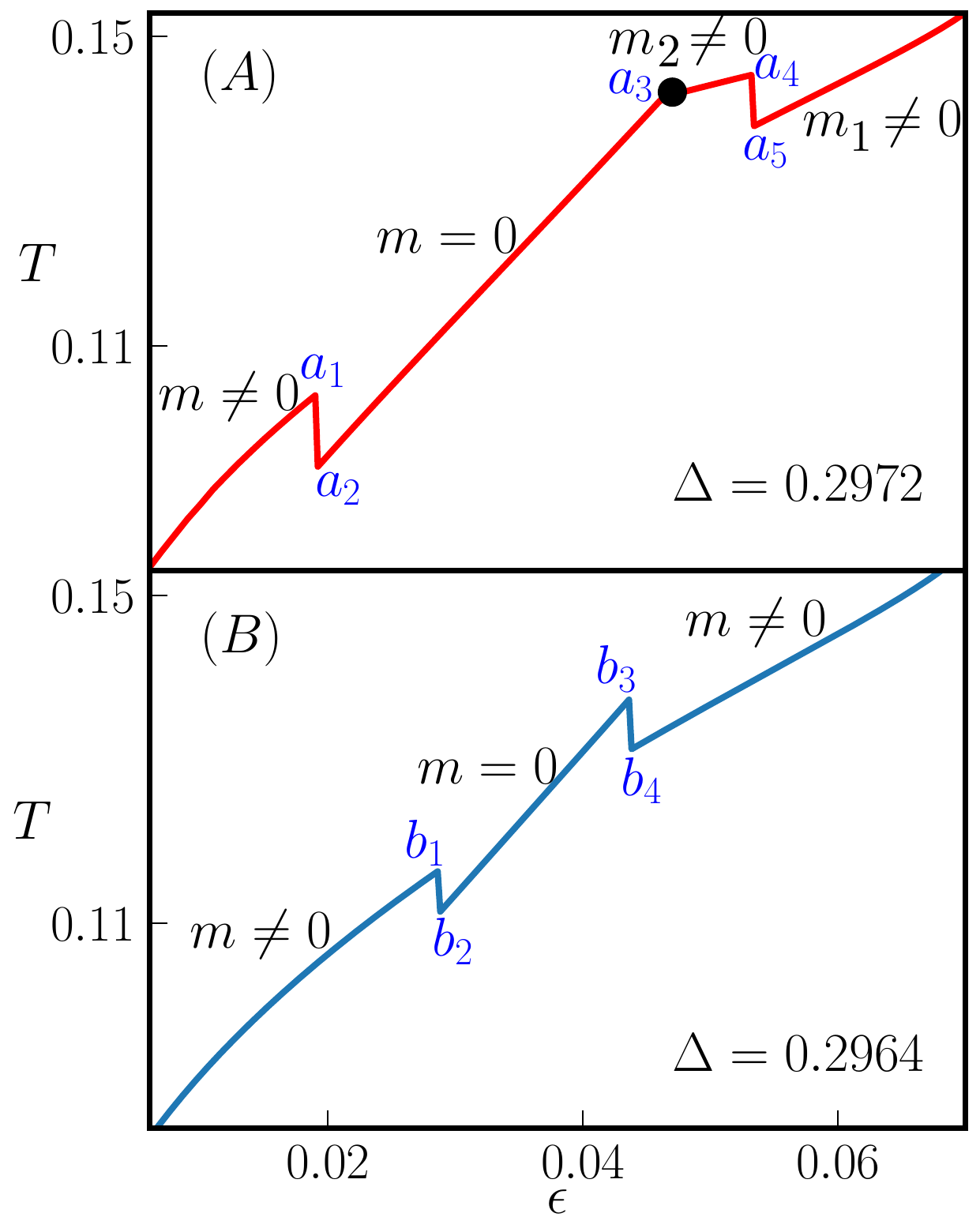}
\caption{The temperature profile (caloric curve) for fixed values of $K=-0.4$ for (A) $\Delta=0.2972$ 
and (B) $\Delta=0.2964$ as a function of $\epsilon$. In panel (A), the discontinuities marked by $a_1a_2$ and $a_4a_5$
correspond to first order transitions and $a_3$ to the continuous transition point. The labels $m_1$ and $m_2$ correspond to two values of the magnetisation order parameter. In panel (B), the discontinuities $b_1b_2$ and
$b_3b_4$ correspond to first order transitions. Equivalent points  in the $T-\Delta$ phase diagram [\fref{fig05}~(B)] are marked by
the same label for comparison.}
\label{fig06}
\end{figure}
\end{center}

\section{Microcanonical phase diagram near the 4th order point}
\label{kplane}

In this section, we consider the microcanonical phase diagram near to the fourth order point. In particular we discuss the phase
diagram for $K=-0.4< K^*_2$ first in the $(\Delta,\epsilon)$ plane, and then in the $( \Delta,T)$ plane. The detailed
$( \Delta,\epsilon)$ phase diagram is plotted in \fref{fig04}. It shows a first order line which extends into the magnetically ordered 
phase and a critical line terminating at a critical end point, as discussed in the preceding section. The inset of \fref{fig04}, which
zooms onto the region where the two lines meet, shows that the first order line curves backward, resulting in re-entrant transitions
as the energy is increased for some narrow range of $\Delta$. This will result in some interesting features of the phase diagram when
plotted in the $(\Delta, T)$ plane.
 
In order to compare the phase diagrams of the two ensembles we now re-plot the $( \Delta,\epsilon)$ phase diagram of \fref{fig04} in
the $(\Delta,T)$ plane. Since some of the interesting features of the phase diagram show up in a rather narrow range of the parameters
we first plot in \fref{fig05}~(A) a schematic phase diagram on a broader scale. A zoomed in non-schematic plot focused on the more
interesting region of the phase diagram is given in \fref{fig05}~(B). In the microcanonical ensemble, a first order transition is characterized by a temperature discontinuity. Thus in the figure one notices that the first order transition is represented by two
lines which give the two temperature values at the transition.  

To get some insight into the phase diagram it is convenient to consider the caloric curve, and plot the temperature as a function of
$\epsilon$ at fixed $\Delta$. This is done for two representative values of $\Delta$: $(i)$ $\Delta=0.2972$, for which two first order
transitions and one second order transition take place, and $(ii)$ $\Delta=0.2964$, where the second order transition is absent. The
first order transitions have to do with the curved (re-entrant) shape of the first order line in the $(\Delta,T)$ plane. The
plots, \fref{fig06}~(A) and (B) show $T(\epsilon)$ for the two respective values of $\Delta$.

\begin{center}
\begin{figure}[]
\includegraphics[width=0.47\textwidth, angle=0]{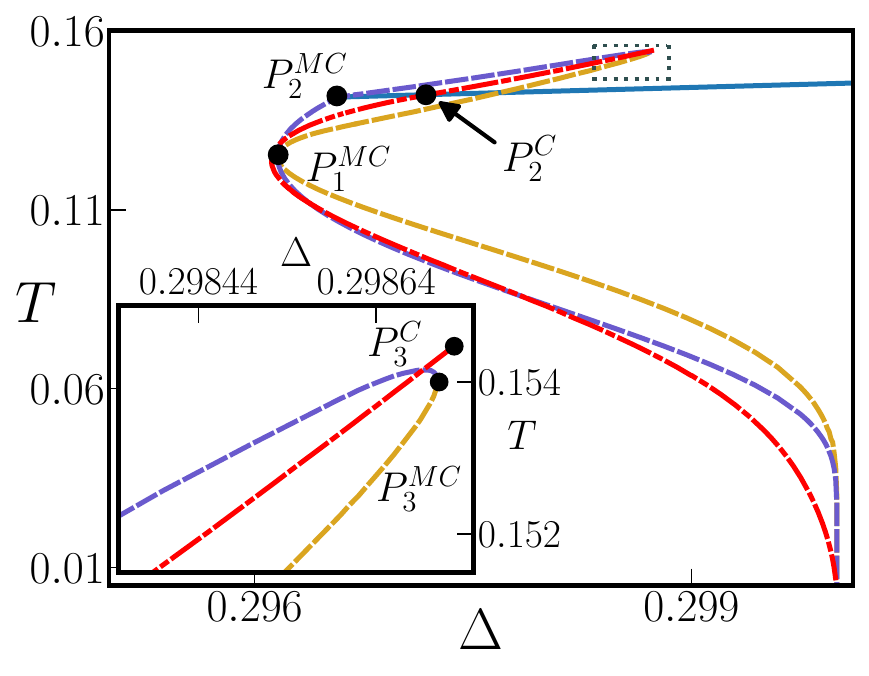}
\caption{(color online) The canonical and microcanonical $(\Delta, T)$ for $K=-0.4<K^*_2$ superimposed. The canonical first order line is shown as
dot-dashed (red) while the corresponding microcanonical transition is represented by dashed lines. The continuous transition
line~(solid blue line) terminates at $P_2^C$ (canonical) and at $P_2^{MC}$ (microcanonical) in the two ensembles. The inset shows
the enlarged version of the region~(dotted square in the main plot) where the first order transition terminates within the ordered
phase at a critical point. In the inset the micro canonical and canonical critical points are labeled as $P_3^{MC}$ and
$P_3^C$~ respectively.}
\label{fig05a}
\end{figure}
\end{center}

\begin{center}
\begin{figure}[t]
\includegraphics[width=0.49\textwidth, angle=0]{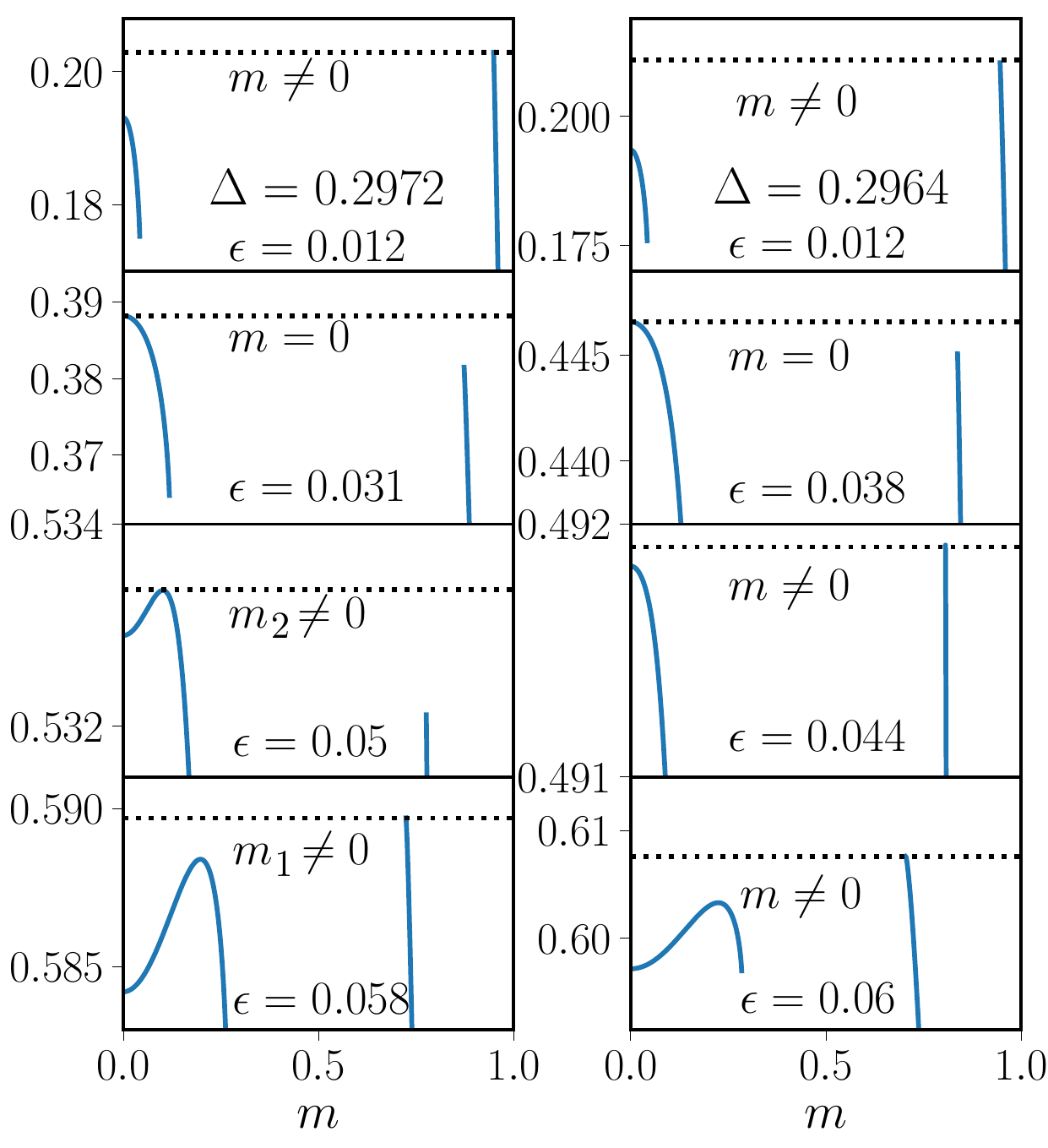}
\caption{The entropy  function $\tilde s_+(\epsilon,m)$  plotted as a function of $m$ for $K=-0.4$  and for $\Delta=0.2972$~(left panels)
and $\Delta=0.2964$~(right panels), for different values of $\epsilon$. In the left [right] panels each plot represents the profile
of $\tilde s_+(\epsilon,m)$ in the different phases highlighted in \fref{fig06}(A)[(B)] as one varies $\epsilon$ at fixed values of
$\Delta=0.2972~~[\Delta=0.2964]$. In the plots, the maxima of the finite $m$ branch (the branch on the right) in
$\tilde s_+(\epsilon,m)$ are not visible, but can be seen as we zoom in.}
\label{fig09}
\end{figure}
\end{center}
  
Consider first \fref{fig06}~(A). At low energy (and low temperature) the $T(\epsilon)$ curve corresponds to a magnetically ordered state.
As the energy increases the temperature undergoes a first order transition into a paramagnetic phase in which the temperature drops
discontinuously from $a_1$ to $a_2$. At a higher value of the energy a second order transition takes place at $T=a_3$, where the system
becomes magnetically ordered again. By increasing the energy even further, the other first order transition into a magnetically ordered
state with a different magnetization is reached, in which the temperature drops from $T=a_4$ to $T=a_5$.  In \fref{fig06}~B the
corresponding behavior for the lower value of $\Delta$ is displayed. Here no second order transition takes place, and there are two first
order transitions: one is a transition from the magnetically ordered state to the paramagnetic state at a low temperature, followed by
a re-entrant transition from the paramagnetic state to the magnetically ordered one at a higher temperature. The corresponding
temperature drops are from $b_1$ to $b_2$ and from $b_3$  to $b_4$, respectively. By considering similar curves at other values
of $\Delta$ one finds that for $\Delta$ corresponding to the point $P_1^{MC}$~$(\Delta\approx0.2961,~T\approx0.1244)$ in \fref{fig05}~B
the two first order transitions merge into a single continuous transition where no discontinuity takes place. At a higher value of
$\Delta$, corresponding to that of $P_3^{MC}$~($\Delta\approx0.29871,~T\approx0.1540$) in \fref{fig05}~B, the first order transition
between the two ordered phases terminates at a critical point, and no such transition exists at higher values of $\Delta$.

To compare the canonical and microcanonical phase diagrams, we superimpose in \fref{fig05a} the two $(\Delta, T)$ phase diagrams for
$K=-0.4$. As is clear from the figure the canonical continuous transition line from the disordered phase coincide with the
microcanonical one. The first order lines separating the disordered and the ordered phases are different in the two ensembles but they
remain close to each other. The critical end points in the two phase diagrams are distinct, with
$(\Delta\approx 0.2966,~T \approx 0.1419)$ at the canonical point $P_2^{C}$, and $(\Delta \approx 0.2972,~T\approx 0.1412)$ at the
microcanonical one $P_2^{MC}$. The two points are very close to each other. Note that in the microcanonical case the first order line
exhibits a discontinuity in its slope at the critical end point, a feature which is absent in the canonical line, whose slope is
continuous at the corresponding critical end point. The zoomed in plot on the transition lines within the ordered phase shows that the
canonical first order line terminates at a critical point $P_3^C$  with $(\Delta\approx 0.29873,~T \approx 0.1544)$ which is close by,
but distinct from the microcanonical one $P_3^{MC}$ located at $(\Delta\approx 0.29871,~T \approx 0.1540)$.

The different phases that we observe along the caloric curve~[see \fref{fig06}], correspond to the values of $m$ at which the entropy
function $\tilde s_+(\epsilon,m)$ is maximized. This is illustrated by plotting $\tilde s_+(\epsilon,m)$ as a function of $m$
for given values of $\Delta$ as $\epsilon$ is changed. In \fref{fig09} we show the plots for $\tilde s_+(\epsilon,m)$ for the same values
of $\Delta$ considered earlier. Due to symmetry between $m$ and $-m$, it is sufficient to consider the positive 
domain of $m$. The left panels of \fref{fig09} are for $\Delta=0.2972$. As also denoted in \fref{fig06}~(A), for
small values of $\epsilon$, $\tilde s_+(\epsilon,m)$ maximizes at non-zero $m$. At higher values of $\epsilon$ the maximum changes
discontinuously to $m=0$~[indicated by the discontinuity in $T$ in \fref{fig06}~(A)] corresponding to the intermediate disordered phase.
As $\epsilon$ is increased to larger values, a continuous transition to an ordered phase~(denoted by $m_2$) takes place, followed by a first order
transition to a different magnetically ordered phase (denoted by $m_1$). A closely similar profile is observed for $\Delta=0.2964$ as seen in the right
panels of \fref{fig09}. However, it lacks the continuous transition for intermediate values of $\epsilon$ but only displays two first
order transitions [Also shown in \fref{fig06}~B)]. 

\section{Conclusions}
\label{conclusions}
In this paper we studied the microcanonical phase diagram of the infinite range Blume-Emery-Griffiths model for negative bi-quadratic
exchange $K<0$, where the canonical phase diagram has been shown to exhibit a fourth order critical point. Studying the phase diagram
of a model near its higher order critical point is of particular interest since, as usual, each type of high order critical point
displays distinct characteristic features of the phase diagram around it. These features tend to persist in quite a broad range of
the model parameter space. The study of the high order critical point of a model thus provide valuable information on its global
phase diagram.

We find that like the canonical phase diagram, the microcanonical phase diagram exhibits a fourth order critical point at different
coordinates ($T, \Delta, K$) compared with the canonical one. This enables one to compare the two phase
diagrams around this point, as is seen in \fref{fig010} and \fref{fig05}. In the vicinity of the microcanonical fourth order point
the transition from the paramagnetic to the ferromagnetic phase can be either continuous or first order. The first order transition
extends into the ferromagnetic phase, thus separating two different magnetically ordered phases. This transition surface is curved
and leads to re-entrant transitions as the energy is varied keeping the parameters of the model $(\Delta, K)$ fixed.
For example, depending on these parameters, as one increases the energy, one may find a sequence of three phase transitions: a first
order transition from $m\ne0$ to $m=0$, followed  by a continuous transition to a phase with $m\ne 0$ and then followed by another
transition separating two magnetically ordered phases. For certain other parameter values the continuous transition is absent and one
encounters a sequence of two first order transitions. At the first order transitions the temperature changes discontinuously. This
rich phase diagram is quite different from its canonical counterpart, including the presence of singular points of first order
transition without a temperature discontinuity.

The difference in the location of the fourth order critical point between the two ensembles, in particular with
$K_2^*\approx-0.0828$ in the microcanonical case and $K^* \approx -0.1838$ in the canonical case, has the consequence that
the $(\Delta,T)$ (or $(\Delta,\epsilon)$) phase diagram for a $K$ value between $K^*$ and $K_2^*$ presents a tricritical point
in the canonical ensemble, while it has a critical end point, together with two different magnetically ordered phases,
in the microcanonical ensemble. This is another marked manifestation of ensemble inequivalence.

A closely related model to (\ref{hamiltonian}) has been studied in \cite{Hoston:1991} where the BEG model with nearest neighbor
couplings (both $J$ and $K$) has been considered within the mean-field approximation in the canonical ensemble. While for positive
bi-quadratic exchange $K>0$ the model is equivalent to the model considered in the present study and yields the same phase diagram 
as that of (\ref{hamiltonian}), for $K<0$ the model exhibits other types of order besides the ferromagnetic one. In particular, for 
negative and large $K$, other phases with ferrimagnetic or antiquadrupolar order have been observed. The phase diagram in this domain 
becomes rather complex with a variety of transitions between the different ordered phases. It would be of interest to extend the present 
study of the microcanonical phase diagram in the large and negative $K$ regime of the model studied in \cite{Hoston:1991} and compare 
it with the canonical one.

\section*{Acknowledgments}
We thank N. Defenu for discussions.  Support by a research grant from the Center for Scientific Excellence at the 
Weizmann Institute of Science is gratefully acknowledged. AC acknowledges financial support from INFN (Istituto Nazionale 
di Fisica Nucleare) through the projects DYNSYSMATH and ENESMA. We thank C. Vanoni for pointing out the misprint in Eqs.~(30)~[Eqs.~(28) in the earlier version].

\end{document}